\documentclass[authoryear,12pt,3p]{jowarticle}

\usepackage[english]{babel}
\usepackage[utf8]{inputenc}
\usepackage{amsmath,amsfonts,amssymb}
\usepackage{graphicx}
\usepackage[colorinlistoftodos]{todonotes}
\usepackage{natbib}
\usepackage{blindtext}
\usepackage{subcaption}
\usepackage{caption}
\usepackage{setspace}
\usepackage{tikz-cd}

\makeatletter
\renewcommand\section{\@startsection{section}{1}{\z@}{-3.25ex plus -1ex minus -.2ex}{1.5ex plus .2ex}{\normalsize\bf}}
\renewcommand\subsection{\@startsection{subsection}{2}{\z@}{-3.25ex plus -1ex minus -.2ex}{1.5ex plus .2ex}{\normalsize\bf}}
\renewcommand\subsubsection{\@startsection{subsubsection}{3}{\z@}{-3.25ex plus -1ex minus -.2ex}{1.5ex plus .2ex}{\normalsize\bf}}
\makeatother

\newtheorem{thm}{Theorem}

\newtheorem{prop}[thm]{Proposition}

\begin{document}

\begin{frontmatter}
\title{On (Some) Gauge Theories of Gravity}
\author{James Owen Weatherall}\ead{james.owen.weatherall@uci.edu}
\address{Department of Logic and Philosophy of Science \\ University of California, Irvine}

\date{\today}

\begin{abstract}
I consider the sense in which teleparallel gravity and symmetric teleparallel gravity may be understood as gauge theories of gravity.  I first argue that both theories have surplus structure.  I then consider the relationship between Yang-Mills theory and Poincar\'e Gauge Theory and argue that though these use similar formalisms, there are subtle disanalogies in their interpretation.
\end{abstract}
\end{frontmatter}

\doublespacing
\section{Introduction} \label{sec:intro}

Motivated by a range of considerations, such as the conceptual conflict between general relativity (GR) and quantum theory \citep{Crowther}, the observation that GR ``breaks down'' in certain domains \citep{EarmanBCWS,WeatherallBreakdown}, and the surprising cosmological observations associated with dark matter and dark energy \citep{Martens+King,Smeenk+WeatherallMG}, many physicists have been drawn to the idea that GR should be ``modified''.  In some cases, these ``modified gravity'' scenarios have been characterized as ``gauge theories of gravity''.  In part this is a marketing scheme: gauge theory -- that is, Yang-Mills theory (YMT), which underlies the Standard Model of Particle Physics -- has many virtues, such as being quantizable (at least in its linearized, weak-field regime) and pertubatively renormalizable.  Authors of these gauge theories of gravity apparently hope to benefit by association.  Even so, it is not entirely clear what the marketers intend to claim.  In particular, as \citet{WeatherallUG} has emphasized, there are multiple distinct ways in which the term ``gauge'' is used in physics.  For instance, ``gauge theory'' is sometimes used to refer to theories with ``surplus'' or ``superfluous'' structure \citep{Redhead, Ismael+vanFraassen, Earman, Healey}; and it is sometimes used to refer to theories with a certain formal structure, viz., ``any physical theory of a dynamic variable which [sic], at the classical level, may be identified with a connection on a principal bundle'' \citep[p. 306]{Trautman}.\footnote{See \citet{Bleecker} \citet{Palais}, \citet{WeatherallFBYMGR}, and \citet{Gomes} for background on theories of this form and their interpretation.  There are plausibly yet other uses of ``gauge theory'', such as theories with primary first-class constraints in the constrained Hamiltonian formalism \citep{EarmanCH}.}

In the present paper, I will investigate these issues within the context of the ``geometric trinity'' of gravity, as introduced by \citet{Trinity}.  As I describe in detail in section \ref{sec:trinity}, the geometric trinity consists of three theories of gravitation that describe the same physical phenomena using different geometrical structures: curvature, as in GR; torsion, as in teleparallel gravity (TEGR) \citep{TPGBook}; and non-metricity, as in symmetric teleparallel gravity (STEGR) \citep{Nester+Chin}.\footnote{We adopt the acronyms used by \citet{Trinity}. Teleparallel gravity in particular has recently been a topic of active research in philosophy of physics: see \citet{Knox}, \citet{Read+TehCQG}, \citet{Meskhidze+Weatherall}, \citet{March+etal}, \citet{Weatherall+Meskhidze}, \citet{Lu+etal}, and \citet{Read+Mulder}. The geometric trinity has been discussed by \citet{Wolf+etalUnderdetermination}, \citet{Wolf+etalNRGT}, and \citet{March+etalTrinity}.}  Each of these are often described as ``gauge theories''; my goal here is to clarify the senses of ``gauge'' on which these claims come out as true.  Previous work has shown that GR is a gauge theory in the second sense \citep{Trautman, WeatherallFBYMGR}, but arguably not the first \citep{WeatherallUG}, and so I will focus on TEGR and STEGR.  In section \ref{sec:structure}, building on previous work by \citet{Weatherall+Meskhidze} and \citet{Lu+etal}, I will show that both of these are gauge theories in the first sense, i.e., they have surplus structure.  Of course, this need not mean they are not also gauge theories in the second sense, and TEGR in particular has been described as both a ``gauge theory of the translation group'' and as a type of ``Poincar\'e Gauge Theory'' (PGT), in analogy to YMT.\footnote{\citet{Lyre+Eynck} offer an early philosophical analysis of both PGT and TEGR, understood as a gauge theory of translation.}  In section \ref{sec:PGT}, I will argue that although TEGR can be presented using similar tools as YMT, its interpretation is importantly different. Section \ref{sec:conclusion} concludes.

\section{The Geometric Trinity}\label{sec:trinity}


We now introduce the geometric trinity. Let $M$ be a smooth, four-dimensional manifold, which we assume to be Hausdorff, paracompact, and parallelizable.\footnote{For background on the formalism used here, using the same notation, see \citet{Wald} or \citet{MalamentGR}.  Parallelizability is a non-standard requirement in GR, but it is necessary and sufficient for the existence of a (global) flat derivative operator.}  For all three of the theories we presently consider, we assume that $M$ is equipped with a smooth, Lorentz-signature metric, which determines the causal structure of spacetime.  Though dynamics will play no role in what follows, in all three theories the metric is dynamically related to the distribution of matter in space and time. Finally, in all three theories, the manifold is also equipped with a (smooth) covariant derivative operator $\nabla$.  The differences between the theories correspond to different choices of derivative operator, for some fixed distribution of matter.  These different choices of $\nabla$ lead to different descriptions of gravitational phenomena.

Recall that in GR, $\nabla$ is taken to be the Levi-Civita derivative operator, i.e., the unique covariant derivative operator on $M$ with the properties that (a) for any smooth scalar field $\alpha$, $\nabla_{[a}\nabla_{b]}\alpha = \mathbf{0}$ and (b) $\nabla_a g_{bc}=\mathbf{0}$.  (When a derivative operator satisfies (a), we say it is \emph{torsion-free}; and if it satisfies (b) we say it is \emph{metric compatible}.)  In general, this derivative operator will have curvature, i.e., it will have non-zero Riemann curvature tensor $R^a{}_{bcd}$, defined as the unique tensor field such that for any smooth vector field $\xi^a$,
\[
R^a{}_{bcd}\xi^b = -2\nabla_{[c}\nabla_{d]}\xi^a.
\]
Curvature represents a failure of path-independence of parallel transport, even locally, for vectors around closed curves.  If curvature vanishes, the derivative operator is said to be \emph{flat}.  We can think of the curvature of a derivative operator as measuring a failure of \emph{integrability}, in the following sense: given a basis for the tangent space at a point $p$, one can extend that basis, by parallel transport, to constant vector fields on a neighborhood of the point just in case curvature vanishes.\footnote{Why think of this as ``integrability''?  There are two ways to think of it.  One is that the equation determining parallel transport is a first order ordinary differential equation, which we integrate to find a solution.  Flatness implies that the integrals performed over different curves agree.  The other, more geometric interpretation involves observing that a derivative operator gives rise to a field of horizontal subspaces of the tangent space to the tangent bundle, and these horizontal subspaces can be locally integrated to be tangent to smooth surfaces if and only if the derivative operator is flat.}

It is standard to characterize models of GR as pairs, $(M,g_{ab})$.  For present purposes, it is useful to expand these somewhat, to include explicit reference to the Levi-Civita derivative operator: $(M, g_{ab}, \nabla)$.  Doing so changes nothing, since $\nabla$ is uniquely determined by $g_{ab}$. We simply make implicit structure explicit.  (This will not be the case for the other theories we consider.) Free massive test bodies in GR follow (timelike) geodesics of $\nabla$, and light rays follow null geodesics.  We take the empirical content of a model to be exhausted by the structure $(M,g_{ab})$, since $g_{ab}$ uniquely determines $\nabla$; and $g_{ab}$ is determined, up to a constant, by the timelike and null geodesics \citep{Weyl,MalamentGR}.

Now consider TEGR.  In this theory, we place different constraints on the derivative operator $\nabla$.  We still assume that $\nabla$ is metric compatible, but now we require it to be flat.  To do this, we must relax the assumption that it is torsion-free, because in general, a derivative operator cannot be metric compatible, torsion-free, and flat, except for very special metrics.  Instead, the derivative operator will have a non-zero torsion tensor, defined so that for an smooth scalar field $\alpha$,
\[T^a{}_{bc}\nabla_a\alpha = 2\nabla_{[b}\nabla_{c]}\alpha.\]
Torsion is a measure of the asymmetry of a derivative operator.  Note that whereas there is a unique torsion-free derivative operator compatible with any metric $g_{ab}$, no analogous result holds for flat, metric-compatible ones once we drop the torsion-free condition.  In other words, given a metric $g_{ab}$, there are \emph{many} flat derivative operators compatible with it.  Torsion, too, can be thought of as measuring a failure of integrability, but now in a different sense.  Since the derivative operator is flat, it also admits cobases of constant covector fields, at least locally.  But those covector fields can be integrated to determine smooth surfaces to which they are normal only if the derivative operator is torsion-free.


As with GR, we can characterize models of TEGR as ordered triples $(M,g_{ab},\tilde{\nabla})$, where $\tilde{\nabla}$ is metric-compatible and flat.  On this theory, free massive test particles no longer follow timelike (affine) geodesics.  Instead, they accelerate according to a force law:
\[
\xi^n\tilde{\nabla}_n\xi^a = -K^a{}_{bc}\xi^a\xi^b,
\]
where $\xi^a$ is the unit tangent vector to the particle's worldline and $K^a{}_{bc}=\frac{1}{2} (T^a{}_{bc}-T_{cb}{}^a - T_{bc}{}^a)$ is the \emph{contorsion tensor} associated with $\tilde{\nabla}$.  Note that these curves, while not geodesics of $\tilde{\nabla}$, \emph{do} correspond to the locally extremal curves of the metric $g_{ab}$, and thus the geodesics of its Levi-Civita derivative operator.  For this reason, we will once again take the empirical content of a model to be exhausted by the structure $(M,g_{ab})$.\footnote{Could torsion have independent empirical significance?  In principle, yes, but in the context of the geometric trinity, it is important that TEGR be taken to be empirically equivalent to GR, since it is supposed to be a mere reformulation.}

Finally, we turn to STEGR.  Here we assume that $\nabla$ is both flat and torsion-free, but we relax metric compatibility.  Instead, we suppose $\nabla$ has a non-zero non-metricity tensor, defined by
\[Q_{abc}=\nabla_ag_{bc}\]
Non-metricity can be thought of as a measure of the degree to which vector lengths and angles, as measured by the metric, are preserved under parallel transport relative to $\nabla$.  Suppose, for instance, that $\xi^a$ is a constant vector field with respect to $\nabla$.  Then it need not follow that the vector have constant length: instead, $\nabla_a(\xi^n\xi^mg_{mn})=\xi^n\xi^mQ_{amn}$.  There are no failures of integrability associated with derivative operators on this theory; there is also no relationship between the metric structure of spacetime and its affine structure.

We can think of models of STEGR as triples $(M,g_{ab},\bar{\nabla})$, where $\bar{\nabla}$ is flat and torsion-free.  Free massive test particles follow accelerated curves, now given by
\[
\xi^n\bar{\nabla}_n\xi^a = -L^a{}_{mn}\xi^m\xi^n
\]
where $L^a{}_{bc}=\frac{1}{2}Q^a{}_{bc}-Q_{(b}{}^a{}_{c)}$ is the \emph{distorsion tensor} associated with $\bar{\nabla}$.  Once again, these curves turn out to be the extremal curves of the metric, and so again we take the empirical content of a model to be exhausted by the structure $(M,g_{ab})$.


\section{TEGR and STEGR have Surplus Structure}\label{sec:structure}

We now turn to assessing the sense whether TEGR and STEGR are gauge theories.  The first thing to say concerns their names: we have been using the acronyms ``TEGR'' and ``STEGR'' throughout the paper, because they are used by \citet{Trinity}.  These stand for ``teleparallel equivalent of GR''and ``symmetric teleparallel equivalent of GR'', respectively.  But in fact, despite the names, these theories are not all (theoretically) equivalent.\footnote{By construction, they are empirically equivalent.}  In particular, \citet{Weatherall+Meskhidze} show that TEGR and GR fail to be categorically equivalent---which in turn means that they are not equivalent by any of the other, generally stronger, criteria that are widely discussed in the literature.\footnote{See \citet{WeatherallTE1,WeatherallTE2} for a review of these criteria of theoretical equivalence. \citet{Lu+etal} have subsequently shown that various other ways of formulating TEGR are categorically equivalent to the presentation \citet{Weatherall+Meskhidze} consider, and all have more structure than GR.}

We will presently show that a similar result holds for STEGR.  But first, it will be helpful to review the result from \citet{Weatherall+Meskhidze}.  The starting point is to define a pair of categories, $\mathbf{GR}$ and $\mathbf{TEGR}$, whose objects are models of the respective theories and whose arrows are ``structure preserving maps''. For convenience, we will consider only isomorphisms.\footnote{Since I present only \emph{inequivalence} results, corresponding results for categories with more arrows whose associated groupoids coincide with the ones we consider will follow immediately.}  In the case of $\mathbf{GR}$, this means models are objects of the form $(M,g_{ab},\nabla)$, defined as above, and the arrows are isometries.  (Note that, since $\nabla$ is a Levi-Civita derivative operator, it is automatically preserved by every isometry.) For TEGR, the objects are structures $(M,g_{ab},\tilde{\nabla})$ and the arrows are isometries that preserve $\tilde{\nabla}$.

Since for both theories, the empirical significance of a model is entirely encoded in the substructure $(M,g_{ab})$, there is a canonical functor relating these theories and preserving empirical content, namely the functor $F:\mathbf{TEGR}\rightarrow\mathbf{GR}$ taking models $(M,g_{ab},\tilde{\nabla})$ to $(M,g_{ab},\nabla)$ and taking arrows to themselves.  If these theories were categorically equivalent, in the sense of \citet{WeatherallNGE}, that functor would be (one half of) an equivalence of categories. But \citet{Weatherall+Meskhidze} show it is not an equivalence.  The reason is that it fails to be \emph{full}: that is, the induced mapping on sets of arrows between pairs of objects of $\mathbf{TEGR}$ fails to be surjective.  This is because not every isometry preserves arbitrary flat, metric-compatible derivative operators.

We now turn to STEGR.  Nearly identical arguments establish that STEGR and GR are also inequivalent.  We work with the same category $\mathbf{GR}$ as before, and we define a new category $\mathbf{STEGR}$ in the obvious way: the objects are structures $(M,g_{ab},\bar{\nabla})$ and the arrows are isometries that preserve $\bar{\nabla}$.  Once again, there is a canonical functor $F':\mathbf{STEGR}\rightarrow \mathbf{GR}$ that takes models $(M,g_{ab},\bar{\nabla})$ to $(M,g_{ab},\nabla)$ and isometries to themselves.  If these were categorically equivalent, $F'$ would be an equivalence of categories.  But it is not.  We have the following result.
\begin{prop}The functor $F':\mathbf{STEGR}\rightarrow\mathbf{GR}$ fails to be full.\end{prop}
Proof. Fix a model $(M,g_{ab},\nabla)$ of $\mathbf{GR}$.  For simplicity, we assume $M$ admits global coordinate systems and that $(M,g_{ab},\nabla)$ has no automorphisms.  Let $(M,g_{ab},\bar{\nabla})$ and $(M,g_{ab},\bar{\nabla}')$ be distinct, non-isomorphic objects of $\mathbf{STEGR}$ corresponding to the same object of $\mathbf{GR}$.  (These always exist. Pick two coordinate systems with distinct coordinate derivatives, and take the models of $\mathbf{STEGR}$ to have those derivative operators.)  Then there is no arrow between $(M,g_{ab},\bar{\nabla})$ and $(M,g_{ab},\bar{\nabla}')$, even though both map to the same model of $\mathbf{GR}$ under $F'$.  It follows that $F'$ is not full, because no arrow between those objects maps to the identity on $(M,g_{ab},\nabla)$.\hspace{.2in}$\square$.

How do these results bear on the sense in which TEGR and STEGR are gauge theories?  \citet{WeatherallUG}, following \citet{Baez+etal}, classifies functors according to what they ``forget''.  An equivalence of categories forgets nothing; any functor that fails to be an equivalence forgets something.  In particular, when a functor fails to be full, one should think of it as ``forgetting structure''.  This is because it shows how the objects in the codomain category have more symmetries than those in the domain category, at least relative to that functor.  In the present case, we see that both TEGR and STEGR involve making additional choices, above and beyond specifying a metric structure.  Arrows of the corresponding categories need to preserve those choices---which is why there are fewer of them than the corresponding models of GR.  This is the sense in which those theories have more structure.

Weatherall goes on to propose that we should say one theory has \emph{surplus structure} just in case there exists another theory, empirically equivalent to the first, and a functor relating their categories of models that preserves the empirical content of those models while forgetting structure.  In that case, one should say that there is an alternative theory that can represent the same phenomena, but which does so with less structure.  Hence, whatever structure distinguishes the theories from one another is unnecessary for representational purposes.  Thus, we have a precise sense in which TEGR and STEGR are gauge theories, in the first sense given above.


\section{Einstein-Cartan vs Yang-Mills}\label{sec:PGT}

We have now seen that TEGR and STEGR both have surplus structure, relative to GR, and thus they are both gauge theories in the surplus structure sense.  But what about the other sense of gauge?  We will not consider STEGR further, as it is relatively uncommon to present that theory as analogous to Yang-Mills theory.\footnote{Uncommon, but not without precedent: see \citep{Adak}.}  But many authors working on TEGR have claimed that it should be understood as either a ``gauge theory of translations'' \citep{TPGBook} or a special case of \emph{Poincare Gauge Theory} (PGT) \citep{Hehl+etal,Hehl,Blagojevic+Vasilic}.  Advocates for both approaches claim that TEGR can be see as a theory of a dynamical connection on a principal bundle.  To the extent that they are correct, TEGR apparently meets Trautman's criterion to be a gauge theory.  But what should we make of these claims?  I will now argue that despite apparent similarities, there are important disanalogies between TEGR and YMT on both approaches.

It will be helpful to first recall how YMT deploys this formalism.  YMT is a theory (or family of theories) of matter in space and time carrying some property -- a type of \emph{charge} that we will call YM-charge -- that influences how that matter evolves.  More precisely: given a relativistic spacetime $(M,g_{ab})$, states of YM-charged matter are represented by sections of a (typically complex) vector bundle $V\rightarrow E\xrightarrow{\pi} M$, usually equipped with some further structure such as a Hermitian inner product and an orientation, which we stipulate is not tangent to the manifold.\footnote{More precisely, and generalizing away from a single spacetime, states of matter are valued in a \emph{gauge natural bundle}, but not a \emph{natural bundle} \citep{March+Weatherall}.}  This vector bundle carries a linear connection, which figures in the matter field's dynamics much like the Levi-Civita derivative operator does in GR. And also like in GR, this connection may be curved, where that curvature is related to the distribution of charge-current density from all matter carrying the same YM-charge by the Yang-Mills equation.

In what sense is this a theory of a dynamical connection on a principal bundle?  The idea is that any vector bundle can be associated with a principal bundle $G\rightarrow P\xrightarrow{\wp} M$, where $G$, called the ``structure group'' of the bundle, is the symmetry group of the (structured) fibers of the vector bundle.  So, for instance, if the fibers are three dimensional complex vector spaces with a Hermitian inner product and orientation, the structure group $G$ would be $SU(3)$.\footnote{More generally, the fibers might be tensor products of vector spaces, with structure inherited from the factors, in which case the structure group would the symmetry group of the factors.  See \citet{GomesPB} for discussion.}  The linear connection on the vector bundle would then uniquely determine a (dynamical) principal connection on this principal bundle, and vice-versa.

This perspective is especially useful because very often, different species of matter all carry the same YM-charge.  (For instance, quarks, antiquarks, and gluons all carry the strong force, but their states are sections of different vector bundles.)   In that case, the matter may have properties valued in different vector bundles. We can still make sense of them carrying the same kind of charge and participating in the same force, however, because all of them are associated to the same principal bundle, and the principal connection on that bundle determines linear connections on all of these associated vector bundles.  Very often, physicists think of the principal connection as the primary object, with the Yang-Mills equation understood as an equation on the total space of the principal bundle.  This is the sense in which YMT is a theory of a dynamical connection on a principal bundle. 

What about TEGR?  As noted, \citet{TPGBook} argue that TEGR can also be seen as a theory of a dynamical connection on a principal bundle over the spacetime manifold, where now the structure group of the principal bundle is the \emph{translation group} $T^4$ on $\mathbb{R}^4$.  In fact, this proposal is controversial, and it is not clear precisely how the mathematics is supposed to work \citep{Fontanini+etal,LeDelliou+etal,Huguet+etal1,Huguet+etal2}.\footnote{\citet{Pereira+Obhukov} defend the approach against some of the criticisms, but they agree on the salient points for the arguments I will presently make.}  But for present purposes, the details of the approach, and of the dispute, do not matter.  It suffices to observe that any theory with the general structure just described would be strikingly disanalogous to YMT.  To see this, observe that since we are interested in gravitational phenomena, and we expect matter experience gravitational influences to be represented by sections of the tangent bundle (or tensor bundles based thereon), we should expect that for a theory analogous to YMT, that the tangent bundle would be associated to whatever principal bundle carries the dynamical connection under consideration and that the principal connection on that bundle would determine a linear connection on the tangent bundle (or some bundle isomorphic to it) via the standard associated bundle construction.  But this cannot happen if the structure group is the translation group.  Broadly, the issue is that the translation group is not the symmetry group of fibers of the tangent bundle, with or without metric.  More generally, there are no non-trivial representations of the translation group in four dimensions on a four-dimensional vector space, because translations do not preserve the origin.  So whatever else may be true, TEGR, viewed this way, uses the formalism differently from YMT.

To address problems closely related to this, \citet{Fontanini+etal} suggest moving to \emph{(pseudo-)Riemann-Cartan geometry}, which can be thought of as a generalization of the pseudo-Riemannian geometry usually used in GR to a context where we think of a manifold of having a local ``Kleinian geometry'' structure.  This is the geometry of PGT.\footnote{Due to space constraints, we suppress some detail. For more, see \citet{Hehl}.  \citet{Lu+etal} offer a very helpful compact exposition of the key ideas at a higher level of generality.  The idea that TEGR is a special case of PGT predates the recent debates; see, for instance, \citep{Blagojevic+Vasilic}.}  A Riemann-Cartan geometry consists of a smooth manifold $M$, a principal bundle $\mathcal{H}\rightarrow P\xrightarrow{\wp} M$, and a \emph{Cartan connection}, which is a Lie algebra-valued 1-form $\eta^{\mathfrak{A}}{}_{\alpha}$ on $P$ satisfying certain further conditions.\footnote{We adopt the notation presented in the appendices to \citet{WeatherallFBYMGR}, though with the caveats explained in the main text and the next note that this is not a principal connection.}  This Cartan connection is \emph{not} a principal connection, because the Lie algebra in which it takes its values is not the Lie algebra of $\mathcal{H}$.  Instead, it is the Lie algebra of another group, $\mathcal{G}$, which has dimension $dim(M)+dim(\mathcal{H})$ and which contains $\mathcal{H}$ as a closed (normal) subgroup.\footnote{Importantly, in addition to other conditions such as equivariance, the connection is required to satisfy the \emph{Cartan} condition, which states that at each point of $P$, $\eta$ is an isomorphism from the tangent space at that point to $\mathfrak{g}$, the Lie algebra of $\mathcal{G}$.}  The torsor of the quotient group $\mathcal{G}/\mathcal{H}$, which is a homogeneous space of the same dimension as $M$, is the ``local Kleinian geometry'' on which the Riemann-Cartan space is modeled.

We can think of $\eta^{\mathfrak{A}}{}_{\alpha}$ as encoding two kinds of information: a true principal connection $\omega^{\mathfrak{A'}}{}_{\alpha}$, valued in the Lie-algebra $\mathfrak{h}$, and a \emph{solder form} on $P$, which in this case is a one-form valued in $\mathfrak{g}/\mathfrak{h}$ that can be interpreted as identifying elements of the tangent space of $M$ at each point with tangent vectors of the ``model space'', i.e., the local Kleinian geometry $\mathcal{G}/\mathcal{H}$.   Note that $\eta^{\mathfrak{A}}{}_{\alpha}$ can also be viewed as (uniquely determining) a principal connection $\alpha^{\mathfrak{A}}{}_{\alpha}$ on a principal $\mathcal{G}$-bundle $\mathcal{G}\rightarrow Q\rightarrow M$, which we may understand as an associated bundle to $P$.

In PGT, we consider the special case where the structure group $\mathcal{H}$ is the Lorentz group in four dimensions, $SO(1,3)$, and the supergroup $\mathcal{G}$ is the \emph{Poincar\'e group}, $SO(1,3)\ltimes T^4$, where $T^4=\mathbb{R}^4$ is the translation group on Minkowski spacetime. Thus, we can think of PGT as a theory of a (dynamical) principal connection on a principal bundle, or as a theory of a (dynamical) Cartan connection on a (different) principal bundle.  (TEGR arises in the special case where the Cartan connection determines a flat affine connection on the tangent bundle.)

Geometrically, the Riemann-Cartan geometry on which PGT is based is unimpeachable, and I am not aware of any criticisms that theory is mathematically problematic.  But as a matter of interpretation, there are several salient disanalogies with YMT.  First, just as we argued above for the $T^4$-bundle approach, four-dimensional vector spaces do not carry a non-trivial representation of the Poincar\'e group, and so the tangent bundle cannot be thought of as associated to the principal bundle $\mathcal{G}\rightarrow Q\rightarrow M$.  So this bundle cannot play the role of the principal bundle in YMT.  On the other hand, there are faithful representations of $SO(1,3)$ on a four-dimensional vector space, and so, under sensible assumptions about the topology of the bundle $\mathcal{H}\rightarrow P\rightarrow M$, the tangent bundle will be isomorphic to a vector bundle associated to $P$; likewise, the Cartan connection also a linear connection of that associated bundle.  This is much closer to YMT.

But there remains a disanalogy, because the Cartan connection does more than induce a connection on its associated bundles.\footnote{There is another disanalogy between PGT and GR / YMT, which is that a Cartan connection is neither a section of a natural bundle over $M$ nor a gauge natural bundle over $P$ \citep{March+Weatherall}.  Thanks to Eleanor March for observing this.}  It also determines an isomorphism between fibers of the associated vector bundle and fibers of the tangent bundle to $M$, via the induced dynamical solder form; and, via this isomorphism, it induces a metric field on the tangent bundle.  This blocks an interpretation of $P$ as the frame bundle for the tangent bundle, because the frame bundle comes equipped with a canonical solder form.  In other words, the Cartan connection -- and Cartan geometry more generally -- can represent the geometrical structure we use in GR and in TEGR, and it can generalize both to allow a combination of curvature and torsion in an elegant geometrical framework.  But the sort of information it encodes, and the way it encodes it, is subtly different from in YMT.

Of course, this sort of disanalogy does not preclude including PGT (and TEGR) among the gauge theories in the second sense; whether one does so is a matter of semantics, and perhaps Trautman exegesis.\footnote{There are very good reasons to think Trautman \emph{would} have called PGT a gauge theory \citep[p. 305]{Trautman}.}  And of course, there are also disanalogies between GR and PGT, also related to torsion and solder forms \citep{Trautman,WeatherallFBYMGR,March+Weatherall}.  But from a conceptual perspective, the differences between PGT and YMT are both subtle and worth emphasizing, even if one does wish to classify them together as ``gauge theories''.

\section{Conclusion}\label{sec:conclusion}

This paper has considered in what senses one should think of TEGR and STEGR as gauge theories of gravity.  After introducing the theories in more detail, I argued that both TEGR and STEGR should be understood as gauge theories in the sense that they both posit ``surplus structure''.  In this, they should be sharply distinguished from GR, which does not posit surplus structure.  I then considered the sense in which TEGR should be understood as a gauge theory by virtue of being a theory of a connection on principal bundle, as suggested by Trautman.  As I argued, while it is true that TEGR can be seen as a theory of a dynamical connection on a principal bundles, details of the interpretation matter.  The different geometrical structures in each case encode different information about their associated vector bundles.  This does not necessarily mean that TEGR (or PGT) should not count as a gauge theory in the second sense, but it suggests that they are, at very least, distinct subspecies.

\section*{Acknowledgments}
This material is based upon work supported by the National Science Foundation under Grant No. 2419967.  I am grateful to Helen Meskhidze for many discussions on these issues and to my fellow symposiasts, Eleanor Knox, Eleanor March, Ruward Mulder, James Read, and William Wolf, for an excellent discussion of the geometric trinity at the 2024 PSA meeting.

\bibliographystyle{elsarticle-harv}
\bibliography{trinity}

\end{document}